\documentstyle[prl,aps,preprint]{revtex}
\begin{document}
\draft
\title{
Evidence for a Bilayer Quantum Wigner Solid
}
\author{H. C. Manoharan, Y. W. Suen, M. B. Santos, and M. Shayegan}
\address{Department of Electrical Engineering, Princeton University,
Princeton, New Jersey 08544}
\date{16 February 1996}
\maketitle  

\begin{abstract}

As the electronic charge distribution in a wide quantum well is
tuned from a single-layer 
through an interacting bilayer configuration to 
weakly-coupled parallel layers, we observe an insulating phase
concurrently manifesting a dramatic evolution.  The data reveal
that interlayer interactions, playing a crucial role, are able to
stabilize a {\it correlated bilayer} electron insulator, thus 
providing tantalizing evidence of a pinned bilayer Wigner solid
phase crystallizing at total filling factor $\nu$ as large as
0.54  ($\nu > \frac14$ in each layer).

\end{abstract}
\vspace{2cm}


It is possible to fundamentally alter the many-body ground states of a 
two-dimensional electron system (2D ES) at high magnetic fields
($B$) through the introduction of an additional 
degree-of-freedom.  For example, the addition of a spin degree-of-freedom
enables the formation of particular spin-unpolarized fractional quantum Hall
(FQH) states observed at lower $B$ \cite{Willett:PRL:5/2,PRL:SpinTrans}, 
while radically increasing the 
perpendicular spatial
degree-of-freedom leads to a weakening and eventual collapse of the 
FQH effect \cite{Shayegan:PRL:Collapse,He:PRB:Collapse}.  
Here we report magnetotransport measurements on an interacting 
{\it bilayer\/} ES confined
in a wide quantum well.  In this system, the additional layer
degree-of-freedom stabilizes new FQH states such as the even-denominator
incompressible liquids at Landau level filling factors $\nu=\frac12$
\cite{Suen:PRL:1/2ab,Suen:1/2c,Eisenstein:PRL:1/2} and
$\nu=\frac32$ \cite{Suen:1/2c}, 
which have no counterpart in standard single-layer 2D ESs. 
Our data reveal an intriguing interplay between the FQH effect and an 
insulating phase (IP) that displays behavior profoundly different from
any observed in a standard 2D ES.  In this paper we focus on this
IP, which we 
demonstrate evolves into a {\it correlated bilayer} electron insulator
with intralayer and {\it inter}layer interactions holding equal
significance, thus providing unique evidence for a pinned bilayer
Wigner crystal (WC).

The sample, grown by molecular beam epitaxy, consists of a 750 \AA\
GaAs quantum well flanked by Al$_{0.35}$Ga$_{0.65}$As
spacers and Si $\delta$-doped layers.  
When cooled 
to low temperature ($T$) in a dilution refrigerator,
this sample has typical dark density 
$n \simeq 1\times 10^{11}$ cm$^{-2}$ and mobility 
$\mu \simeq 1\times 10^{6}$ cm$^2$/Vs. 
Both $n$ and the charge distribution symmetry are controlled 
via front- and back-side gates \cite{Suen:PRL:1/2ab,Suen:1/2c,Suen:PRB:WQW}.
In the 
density range spanned by our experiments, 
$3.7 \times 10^{10} < n < 19.0 \times 10^{10}$ cm$^{-2}$, at most 
two subbands are occupied. 
As 
electrons are added to the wide quantum well, their electrostatic repulsion
causes them to pile up near the sides of the well, and 
the resulting
electron distribution appears increasingly bilayer-like as $n$ grows
[Fig.\ 1(inset)] \cite{Suen:PRB:WQW}.
For symmetric charge distributions, i.e.\ ``balanced'' states,
two relevant parameters that quantify this evolution are the 
symmetric-to-antisymmetric energy 
gap $\Delta_{\text{SAS}}$, which is
a measure of the coupling
between the two layers,  and the interlayer distance
$d$ [Fig.\ 1(inset)]. 
Experimentally, $\Delta_{\text{SAS}}$ is deduced from Fourier transforms
of the low-$B$ Shubnikov-de Haas oscillations
at various $n$.  We then perform 
self-consistent Hartree-Fock calculations at zero $B$, 
with the well width as a 
single fitting parameter, to match the calculated $\Delta_{\text{SAS}}$
with the measured values.  In general, the agreement between
the calculated and measured $\Delta_{\text{SAS}}$ is excellent
\cite{Suen:PRL:1/2ab,Suen:1/2c,Suen:PRB:WQW}, and
$d$ may then be reliably
deduced from the calculated charge distributions
\cite{HartreeFock}.  
A crucial property of the 
ES in a wide quantum well is that, for a given well width, both 
$\Delta_{\text{SAS}}$ and $d$ depend on $n$: increasing $n$ makes $d$ larger
and $\Delta_{\text{SAS}}$ smaller so that the system can essentially
be tuned from a bilayer ES at high $n$ to a (thick) 
single-layer-like system by decreasing $n$ [Fig.\ 1(inset)].  
This evolution with $n$
plays a critical role in the properties of the correlated 
electron states.

The evolution of the FQH states in this 
system has been studied recently \cite{Suen:1/2c}.
As this evolution provides a
context for understanding the IP, we briefly summarize now the 
overall behavior.
At the lowest
$n$ the data exhibit the usual FQH effect at odd-denominator fillings, 
while at the highest $n$ the strongest FQH states are those with 
{\it even numerators\/}, as expected for a system of two 2D layers in 
parallel.  For intermediate $n$, {\it even-denominator\/} FQH states
at $\nu = \frac12$ and $\frac32$, which are stabilized by 
both interlayer and intralayer correlations,  are observed.  
Concurrent with this evolution of the
FQH states, we observe an IP which moves to {\it higher\/} $\nu$ as
$n$ is increased.  

Before discussing the FQH state evolution in more detail, it is
instructive to first examine the 
spectacular evolution of this IP.  Its behavior is
summarized in Fig.\ 1, where the diagonal
resistivity $\rho_{xx}$ at
base $T$ ($\simeq 25$ mK)
is plotted vs $\nu^{-1} \propto B$ for several representative
$n$.  
Experimentally, the IP is identified by a
$\rho_{xx}$ that is both  
large ($> h/e^2 \simeq 26$ k$\Omega/\Box$,
the quantum unit of resistance) as well
as strongly $T$-dependent.  For very low $n$, the IP appears near 
$\nu = \frac15$ (trace A), while at the highest $n$ there is an IP for
$\nu \lesssim \frac12$.  The IP observed in the intermediate density
range ($10 \times 10^{10} \lesssim n \lesssim 14 \times 10^{10}$ cm$^{-2}$) 
is most remarkable as it very quickly moves to larger $\nu$ with 
small increases in $n$ (see, e.g., traces B, C, and D); 
along the way, 
it also shows {\it reentrant\/} behavior around well-developed
FQH states at $\nu = \frac27$ (trace B), 
$\nu = \frac13$ (traces C and D), and 
$\nu = \frac12$ (bold trace E).  Then, as $n$ increases past this point,
the IP begins to move in the opposite direction to lower $\nu$ (trace F).

The data in Fig.\ 1 for
$n = 12.6 \times 10^{10}$ cm$^{-2}$ are plotted in more
detail in Fig.\ 2, along with the associated Hall resistivity
and the $T$-dependence of both the reentrant IP peak (showing 
a diverging $\rho_{xx}$ as $T\rightarrow 0$) and the
$\nu = \frac12$ minimum (exhibiting activated behavior 
characteristic of a FQH liquid state with finite energy gap).
As a whole, the data of Fig.\ 2 bear a striking resemblance 
to the IP observed reentrant around $\nu = \frac15$ in low-disorder,
single-layer 2D ESs, generally interpreted as a pinned Wigner solid
\cite{Shayegan:LDES:WCreview}; 
here, however, the IP is reentrant around the bilayer
$\nu = \frac12$ FQH state, with the reentrant peak reaching the
prominently high  
filling of $\nu = 0.54$!  

The IPs presented in Fig.\ 1 cannot be explained by single-particle
localization.  First, in the case of standard, single-layer 2D ESs it is
well known that as $n$ is lowered, the quality of the 2D ES deteriorates
and the sample shows a disorder-induced IP at progressively larger
$\nu$ \cite{Sajoto:PRB:LowDens}.  This is opposite the behavior
observed here: as $n$ decreases from 
$19.0 \times 10^{10}$ to $3.7 \times 10^{10}$ cm$^{-2}$,
the quality worsens somewhat as expected
(e.g.\  mobility decreases monotonically
from $1.4 \times 10^{6}$ to 
$5.3 \times 10^{5}$ cm$^2$/Vs) but the 
IP moves to {\it smaller\/} $\nu$.  Second, the observation of IPs which 
are reentrant around {\it correlated\/} FQH states, and particularly around
the very fragile and disorder-sensitive
$\nu = \frac12$ state \cite{Suen:1/2c}, strongly
suggests that electron interactions are also essential
for  stabilizing the IP.

To illustrate that the behavior of this IP is indeed consistent with 
the WC picture, it is important to elaborate on the evolution of 
FQH states in an ES confined in a wide quantum well \cite{Suen:1/2c}. 
This evolution can be understood by 
examining the competition between (1) $\Delta_{\text{SAS}}$, (2) the
in-plane correlation energy $Ce^2/\epsilon\ell_B$ [where
$C$ is a constant $\sim 0.1$ and 
$\ell_B \equiv (\hbar/eB)^{1/2}$ is the magnetic
length], and (3) the interlayer
Coulomb interaction ($\sim e^2/\epsilon d$).  To quantify behavior
it is useful to construct the ratios $\gamma \equiv 
(e^2/\epsilon\ell_B)/\Delta_{\text{SAS}}$ and 
$(e^2/\epsilon\ell_B)/(e^2/\epsilon d) = d/\ell_B$.  As $n$ is increased,
$\gamma$ increases since both $\Delta_{\text{SAS}}$ and $\ell_B$ (for
a FQH state at a given $\nu$) decrease, and $d/\ell_B$ increases. 
Experiments show that when $\gamma$ is small, 
the ES exhibits only ``one-component'' (1C) FQH states (standard
single-layer
odd-denominator states) constructed solely from the symmetric subband,
while for large $\gamma$ the in-plane Coulomb energy becomes
sufficiently strong to allow the antisymmetric subband to mix into the 
correlated ground state to lower its energy, and a ``two-component'' (2C)
state ensues.  These 2C states, constructed out of the 
now nearly degenerate symmetric and antisymmetric basis states, come
in two classes. For large $d/\ell_B$, the ES behaves as two
independent layers in parallel, each with density $n/2$;
FQH states in this regime therefore have even numerator and odd
denominator.  For small
enough $d/\ell_B$, the interlayer interaction can become comparable
to the in-plane interaction and a fundamentally new kind of FQH state
becomes possible.  Such a state has strong interlayer correlation and 
can be at even-denominator $\nu$.  A special example is the
$\Psi_{331}$ state associated with the $\nu = \frac12$  FQH
state observed in bilayer ESs with appropriate
parameters \cite{Suen:PRL:1/2ab,Suen:1/2c,Eisenstein:PRL:1/2}.

We have determined the quasiparticle excitation gaps 
$\Delta_{\nu}$ of several 
FQH states in the current system for several $n$ via thermal activation
measurements; these gaps are plotted vs $\gamma$ in Fig.\ 3(a).
As expected, we observe that increasing $\gamma$ suppresses 1C
states and enhances 2C states.   Two states, $\nu = \frac23$
and $\nu = \frac43$, undergo a 1C to 2C phase transition
as $\gamma$ is increased \cite{Suen:1/2c}, as evidenced by
sharp minima in their gaps.  The critical point for 
this transition, $\gamma \simeq 13.5$, matches the point where
the gaps of other 1C and 2C states emerge from zero. 
Surrounding this point is a region
where the $\nu = \frac12$ FQH liquid stabilizes.  
Note that since this 
2C state also possesses interlayer correlation
(the 2C $\nu = \frac23$ and $\frac43$ states 
are simply $\frac13$ and $\frac23$ states
in parallel layers), it exists only within a finite range of $\gamma$.
The relevance of this plot to the IP is immediately  highlighted by 
examining the three main reentrant peaks in Fig.\ 1 (from traces
B, D, and E), which appear at $\nu = 0.30$, 0.39, and 0.54 for the 
IPs surrounding the $\nu = \frac27$, $\frac13$, and $\frac12$ FQH states,
respectively.  The values of $\gamma$ at these reentrant peaks
are respectively 16.9, 16.3, and 16.5.  The peak 
positions span a large 
range of $\nu$, and yet the associated $\gamma$ are remarkably similar.
Moreover, at this value of $\gamma \simeq 16.5$, interlayer interactions
are clearly important as this point is straddled by the 
2C $\frac12$ state in Fig.\ 3(a).

The construction of a phase diagram for the observed IPs
facilitates a lucid connection between the IP evolution, the 
1C to 2C transition, and the development of the
$\nu = \frac12$ liquid.  To this end, we first collected a
$\rho_{xx}$ dataset for a fairly dense
grid of points in the $n - B$ plane by incrementally changing
$n$ and sweeping $B$ at base $T$.  Next, $\rho_{xx}$
was mapped to a color interpolating 
between blue ($\rho_{xx} = 0$) and 
red ($\rho_{xx} \ge h/e^2$).  Finally, using known values
for $B$, $n$, and $\Delta_{\text{SAS}}$ at each point,
the  color-mapped $\rho_{xx}$ dataset was plotted vs
$\nu$ and $\gamma$ [Fig.\ 3(b)].  
By utilizing $h/e^2$ as a natural
resistivity scale for demarcating the IP and non-insulating
states \cite{Shahar:PRL:Univ}, the result is a comprehensive phase
diagram depicting incompressible phases (dark blue) together with 
compressible phases, both insulating (dark red) and metallic
(all other colors).

Immediately obvious in the phase diagram are the various FQH
transitions, manifested by the appearance or disappearance of 
dark blue FQH phases at several $\nu$ (see, e.g., $\frac35$, 
$\frac45$, and the $\frac12$ ``island''), or by a change in 
vertical width of the FQH phase (see, e.g., $\frac23$); these
transitions correlate directly with the measured energy gaps
[Fig.\ 3(a)].
Another striking feature is the wrinkling
in the IP boundary:  this is caused by the reentrance of the IP around
several FQH states as discussed earlier.  The limiting critical
$\nu$ at
low $n$ is close to $\frac15$, consistent with a low-disorder single-layer
2D ES, while for the highest $n$, where the 
ES is effectively two layers in parallel, the IP is present for
$\nu \lesssim \frac12$, i.e.\ $\nu \lesssim \frac14$ in each layer.
This is reasonable considering that even at the largest $n$, interlayer
interactions are present in this ES as evidenced by the observation
of a {\it correlated\/} $\nu = 1$ QH state at high $n$ in the same
well \cite{Lay:PRB:Nu1}.  Such interactions can move the WC ground state
to $\nu \simeq \frac14$ (for each layer), somewhat larger than 
$\frac15$ expected if there are no interlayer interactions (see below).
We note, however, that our measurements on wider quantum well samples
indicate that in the high-$n$ limit,
the onset of the IP indeed approaches $\nu \simeq \frac25$, consistent
with two high-quality parallel layers becoming insulating near $\frac15$
filling in each layer in the absence of interlayer interactions.
This regime is evidently outside the density limits of our current sample.

We can examine in more detail the evolution of the IP as depicted
in the phase diagram of Fig.\ 3(b) by making comparisons to Fig.\ 3(a).
For intermediate $n$, as $\gamma$ increases,
the IP first remains close to 
$\nu = \frac15$ but then begins to move
to higher $\nu$ in the range $12 < \gamma <15$.  This range
is precisely bisected by 
$\gamma \simeq 13.5$ [Fig.\ 3(a)] where the 1C to 2C
transition occurs.  Then the IP moves very quickly
to $\nu \simeq \frac12$ as evidenced by the nearly vertical phase
boundary at $\gamma \simeq 16$.  As discussed earlier
and as evident from Fig.\ 3,  this $\gamma$ is centrally 
located in the parameter range in which the $\frac12$ state stabilizes.
A quick glance at the phase diagram underscores this central point:
the $\gamma$-extent of the $\nu = \frac12$ island coincides directly
with the rapid $\nu$-shift in the phase boundary of the insulator.

Up to now we have focused exclusively on symmetric 
(``balanced'') charge distributions.
Intuitively, however, similar to the 2C FQH states
\cite{Suen:1/2c},
the strength of a bilayer WC should be diminished under unbalanced
conditions.	This is indeed observed quite prominently
in our system, and can be highlighted by examining the high-$\nu$
reentrant IPs. 
Figure 4 shows the effect of asymmetry on the IP reentrant
around $\nu = \frac13$ at fixed 
$n = 11.0 \times 10^{10}$ cm$^{-2}$ and for varying $n_t$,
where  $n_t$ is the electron density transferred from the back 
layer to the front by proper gate biasing
from the balanced condition.  It is very clear that,
while the 1C $\nu = \frac 13$ state is strengthened as expected,
the IP is weakened
by increasing imbalance $|n_t|$: the IP is most stable in a perfectly
balanced state, and the IP peak at $\nu = 0.38$ drops dramatically
even for a small imbalance $n_t = 4.6 \times 10^{9}$ cm$^{-2}$.
As $n_t$ is increased past $\simeq 7 \times 10^{9}$ cm$^{-2}$,
the reentrant IP is destroyed ($\rho_{xx} < h/e^2$)
and the $\frac 12$ state disappears.
For the IP reentrant around $\nu = \frac12$ at 
$n = 12.6 \times 10^{10}$ cm$^{-2}$, 
the corresponding destabilization of the insulator (not shown
here) occurs at an imbalance of less than
3\% ($|n_t|/n \simeq 0.027$). 
In all cases, note that both the $\nu = \frac 12$
state and the reentrant IP are strongest in the balanced condition;
asymmetry simultaneously destroys both the bilayer quantum liquid
{\it and} the insulator.

Finally, 
it is plausible that interlayer interactions can modify the ground-state
energies so that for appropriate parameters a crossing of the liquid
and solid states occurs at the large fillings we identify (e.g.,
$\nu = 0.54$, i.e.\ $\nu > \frac14$ in each layer).  Calculations
\cite{Oji:PRL:BiWC} indicate that the effect of interlayer Coulomb
interaction
can be particularly strong near the magnetoroton minimum and lead
to the vanishing of the FQH liquid gap.  This vanishing can be
associated with an instability toward a ground state in which each
of the layers condenses into a 2D WC \cite{Oji:PRL:BiWC}.

Before concluding, it is beneficial to set this work in context with
previously-reported reduced-dimensional insulators
\cite{Shayegan:LDES:WCreview}.  Our bilayer electron 
system provides a unique means of tuning the effective electron-electron
interactions underpinning the formation  of
various many-particle ground states.  The crux of
this reasoning is that this system possesses two vital ``yardsticks''
for gauging the relative importance of inter- and intralayer interactions:
the 1C to 2C transition and the novel bilayer $\nu = \frac12$
condensate.  Utilizing these unimpeachable measuring sticks,
we can connect the fascinating evolution of the IP with	the significance
and critical counterbalance of electron-electron interactions.  In doing 
so, we believe this work transcends specificity to bilayer systems, and
provides convincing evidence that electron-electron correlations
are a compulsory component of IPs observed in all 2D
systems possessing similarly low disorder.

In summary, our data conclusively indicates that the IP we observe
for $\gamma \gtrsim 13$ 
is a collective 2C state with
comparable interlayer and intralayer correlations.  The 
characteristics of this bilayer
electron insulator are remarkably consistent with the
formation of a novel pinned 
bilayer-correlated Wigner solid, a unique 2D electron
crystal stabilized through the introduction of an additional
quantum degree-of-freedom.

We thank X. Ying, S.~R.~Parihar, D.~Shahar, and L.~W.~Engel  
for technical assistance. H.\ C.\ M.\ is grateful to 
the Fannie and John Hertz Foundation for fellowship support.
This work was supported by the
NSF.  


\begin{figure}
\caption{
Evolution of the IP at
$T \simeq 25$ mK for varying $n$ (specified in units of
$10^{10}$ cm$^{-2}$ within legend).
Inset: Conduction band potentials (solid curves) and electron density
profiles (dotted curves) from self-consistent Hartree-Fock simulations
for increasing $n$.
}
\label{fig1}
\end{figure}

\begin{figure}
\caption{
Diagonal and Hall resistivity vs $B$ at 
$n = 12.6 \times 10^{10}$ cm$^{-2}$, highlighting the 
reentrant IP around the bilayer $\nu = 1/2$ FQH liquid.
Inset: $T$-dependence of the $\nu = 1/2$ minimum
and the reentrant peak at $\nu = 0.54$.
}
\label{fig2}
\end{figure}

\begin{figure}
\caption{
(a) Measured energy gaps $\Delta_\nu$ of several FQH states vs
$\gamma$.  The number of components (C) in each state is shown
in parentheses.  (b) Phase diagram showing $\rho_{xx}$,
chromatically mapped according to the colorbar (right), 
vs $\nu$ and $\gamma$.
}
\label{fig3}
\end{figure}

\begin{figure}
\caption{
Effect of asymmetry on the reentrant IP.  Traces are at fixed
total $n$, with varying amounts of charge $n_t$ transferred between
layers.	 Corresponding traces for negative $n_t$ show equivalent
behavior: slight imbalance $|n_t|$ destabilizes the IP.
}
\label{fig4}
\end{figure}


\begin{references}
\bibitem{HartreeFock} Both $\Delta_{\text{SAS}}$ and $d$ may
change in $B$ by an amount that is 
difficult to determine.  
The zero-$B$ parameters are nevertheless instrumental
in quantifying the high-$B$ behavior.
\end{references}
\end{document}